\documentclass[10pt]{article}
\usepackage[a4paper, total={7in,10in}]{geometry}
\usepackage[utf8]{inputenc}
\usepackage{amsmath}
\usepackage{amsfonts}
\usepackage{amssymb}
\usepackage{graphicx}
\usepackage{hyperref}
\usepackage{authblk}
\hypersetup{
    colorlinks=true,
    citecolor=red
}

\newcommand{\be}{\begin{equation}}
\newcommand{\ee}{\end{equation}}

\begin{document}
\title{Spectral data analysis methods for the two-dimensional diagnostics}
\author[1]{Minjun J. Choi}
\affil[1]{National Fusion Research Institute,  Daejeon 34133, Korea, Republic of}

\maketitle

\abstract{Some spectral data analysis methods that are useful for the two-dimensional diagnostics data are introduced. 
It is shown that the frequency spectrum, the local dispersion relation, the flow shear, and the nonlinear energy transfer rates can be estimated using the proper analysis methods.
}

\section{Introduction}

Various two-dimensional diagnostics have been developed to study the complicated dynamics of the tokamak fusion plasma.
For example, the electron cyclotron emission imaging (ECEI) diagnostics, microwave imaging reflectometry (MIR), beam emission spectroscopy (BES) are used to measure the local electron temperature or density fluctuation and provide a more comprehensible picture of plasma instabilities. 
Most advantages of the two-dimensional diagnostics come from the inherent high-resolution multi-channel characteristics.
However, its strengths would be fully exploited when a proper data analysis method is used. 
There are already well developed spectral analysis methods in the literature.
Correct understanding and application of the methods are critical to draw meaningful results. 
In this paper, we provide some practical examples that show how the proper spectral methods can draw the meaningful information from the two-dimensional data. 
This paper is organized as follows. 
In Section 2 some linear and nonlinear spectral methods will be reviewed, and in Section 3 the practical application examples using the ECEI data are provided.
The frequency spectrum, the spatial derivative of the plasma flow, and the nonlinear energy transfer rates can be accurately estimated using the two-dimensional data.
In Appendix, the Python code package developed for the spectral and statistical analysis of the KSTAR diagnostics data is introduced briefly. 

\section{Linear and nonlinear spectral analysis methods}

\subsection{Frequency spectrum and the coherence estimation}

The frequency spectrum of a discrete time series data is often estimated by the Fourier transform~\cite{Welch:1967tq}. 
Let $X(f_p)$ be the discrete Fourier transform (DFT) of a discrete time series data $x(t_j)$. 
For each frequency index $p$, the DFT coefficient $X(f_p)=X_p$ is a complex value and it can be written as $X_p = A_p e^{i \alpha_p}$ where $A_p$ represents an amplitude of the $f_p$ frequency sinusoidal oscillation in the time series data and $\alpha_p$ is the initial phase. 

\begin{figure}
\includegraphics[keepaspectratio,width=0.25\textwidth]{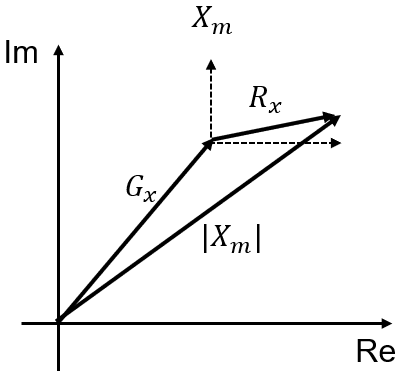}
\centering
\caption{Representation of the Fourier transform coefficient in the complex plane}
\label{Fcxp}
\end{figure}

If an ideal fluctuation with the wavenumber $\bold{k}$ and the amplitude $G$ is measured as an oscillation with the frequency $f=f_m$ by some diagnostics, we may expect to get $|X_m| = G$ from the DFT of the recorded signal. 
However, there is always noise component in the measured signal. 
$X_m$ will include the fluctuation part ($G_x e^{i \delta_x}$) and the noise part ($R_x e^{i n_x}$) where $G_x$ and $R_x$ are the measured fluctuation and noise amplitudes of the frequency $f_m$ in $x(t_j)$, and $\delta_x$ and $n_x$ are their initial phases, respectively (figure~\ref{Fcxp}). 

\begin{figure}
\includegraphics[keepaspectratio,width=0.6\textwidth]{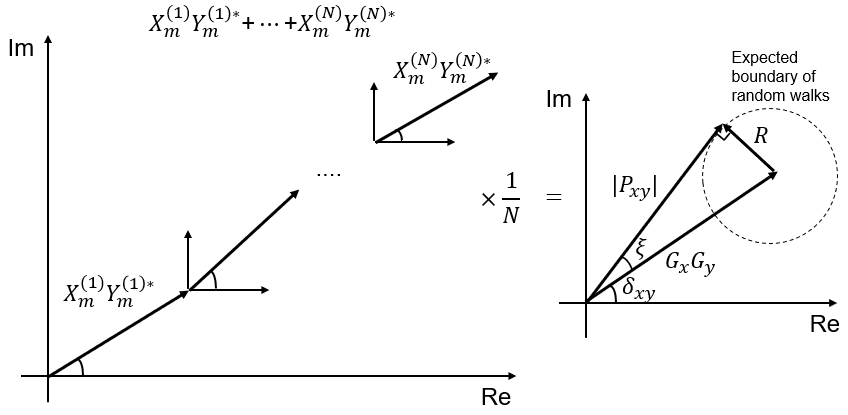}
\centering
\caption{Ensemble average in the complex plane}
\label{Feac}
\end{figure}

To reduce the noise contribution in the frequency spectrum estimation, we can utilize another measurement $y(t_j)$ by an adjacent channel separated by $\bold{d}$ and $|\bold{d}| < \lambda_c$ where $\lambda_c$ is the correlation length of the fluctuation. 
Let $Y(f_p)$ be the DFT of $y(t_j)$. 
At the given frequency $f_m$, $Y_m$ will also be composed of $G_y e^{i \delta_y}$ and $R_y e^{i n_y}$ where $G_y$ and $R_y$ are the measured fluctuation and noise amplitudes in $y(t_j)$, and $\delta_y$ and $n_y$ are their initial phases, respectively. 
Then, it is well known that the cross power ($P_{xy}$) can be used to reduce the noise contribution in the spectrum via the ensemble average of the $N$ independent measurements of $X Y^*$.
\be
P_{xy} = \langle X Y^* \rangle = \frac{X^{(1)} Y^{(1)*} + ... + X^{(N)} Y^{(N)*}}{N} 
\ee
where the superscripts indicate the measurement number. 
At $f_m$, the 1st measurement can be written as follows.
\begin{eqnarray}
X^{(1)}_m Y^{(1)*}_m &=& (G_x^{(1)}  e^{i \delta_x^{(1)}} + R_x^{(1)} e^{i n_x^{(1)}})(G_y^{(1)} e^{-i \delta_y^{(1)}} + R_y^{(1)} e^{-i n_y^{(1)}})  \nonumber \\
&=& G_x^{(1)}G_y^{(1)}  e^{i (\delta_x^{(1)}-\delta_y^{(1)})} + G_x^{(1)} R_y^{(1)}  e^{i (\delta_x^{(1)}-n_y^{(1)})} + R_x^{(1)} G_y^{(1)} e^{i (n_x^{(1)}-\delta_y^{(1)})} + R_x^{(1)} R_y^{(1)} e^{i (n_x^{(1)}-n_y^{(1)})}
\end{eqnarray}
In the above equation, each term can be thought as a vector in the complex plane.
If we choose the separation time between the measurements longer than the correlation time of the noise, the initial phases of the noise ($n_x$ and $n_y$) and all $\delta_x - n_y$, $n_x - \delta_y$, and $n_x - n_y$  will be random in each measurement.
Adding up vectors in the complex plane with the random phase will be like a random walk (RW) diffusion whose expected deviation is $\mathrm{(step~size)}\times \sqrt{N}$.
The division by $N$ in the ensemble average will make the random phase terms decay with $1/\sqrt{N}$. 
On the other hand, the first term has the non-random definite phase, $\delta_x^{(1)}-\delta_y^{(1)}$, which is the ideal fluctuation phase difference between two channels, i.e. $\delta_x^{(1)}-\delta_y^{(1)} \equiv \delta_{xy} = \bold{k} \cdot \bold{d}$. 
The ensemble average in the cross power calculation will lead to  
\be
\langle X_m Y_m^* \rangle = G_x G_y e^{-i \delta_{xy}} + \frac{\mathrm{RW}(G_x R_y) + \mathrm{RW}(G_y R_x) + \mathrm{RW}(R_x R_y)}{N}
\ee
where $\mathrm{RW}(C)$ represents the random walk with the amplitude $C$ in the complex plane (figure~\ref{Feac}). 
The amplitude of the cross power can be written as 
\be
|\langle X_m Y_m^* \rangle| \approx G_x G_y \pm \mathcal{O}\left(\frac{G_x R_y}{\sqrt{N}}\right) \pm \mathcal{O}\left(\frac{G_y R_x}{\sqrt{N}}\right) \pm \mathcal{O}\left(\frac{R_x R_y}{\sqrt{N}}\right) 
\ee
Therefore, the cross power provides a significant noise reduction in the spectrum estimation for the large $N$. 
CORRELATION ECE 
The two-dimensional diagnostics with many channels is favorable to find a proper pair of channels to calculate the cross power for a specific mode or event. 

Note that it is convenient to normalize the cross power spectrum with the auto power spectra as follows.
\be
\gamma_{xy} = \frac{|\langle X Y^* \rangle|}{\sqrt{\langle XX^* \rangle} \sqrt{\langle YY^* \rangle}}
\ee
It is called the coherence and ranges from 0 to 1. 
It means the coherent ($\bold{k} \cdot \bold{d} \approx \mathrm{const}$) fluctuation power fraction against the total power at each frequency.
The noise floor in the coherence will be given as $\frac{|\langle R_x R_y e^{-i(n_x - n_y)} \rangle|}{R_x R_y} \approx 1/\sqrt{N}$. 

\subsection{Local wavenumber and the flow shear estimation}

The phase difference between the measured signal $x(t_j)$ and $y(t_j)$ at different positions also contains the important information. 
This phase difference, or cross phase, can be obtained as 
\be
\theta_{xy} = \tan^{-1}\left[ \frac{\mathrm{Im}[P_{xy}]}{\mathrm{Re}[P_{xy}]} \right]
\ee
where $P_{xy}$ is the cross power. 
From figure~\ref{Feac}, we can see that $\theta_{xy} = \delta_{xy} \pm \mathcal{O}(\xi)$ where $\delta_{xy}$ is the phase difference by the mode $\bold{k}$ between two channels, i.e. $\delta_{xy} = \bold{k} \cdot \bold{d}$, and $\xi$ is the noise contribution.
The maximum possible noise contribution is  
\be
\xi = \sin^{-1} \left[ \frac{R}{G_x G_y} \right] \approx \frac{2 \epsilon}{\sqrt{N}} + \frac{\epsilon^2}{\sqrt{N}} + \mathcal{O}\left(\frac{\epsilon^3}{N^{3/2}}\right ) + ...
\ee
where $R = \frac{G_x R_y}{\sqrt{N}} + \frac{G_y R_x}{\sqrt{N}} + \frac{R_x R_y}{\sqrt{N}} $ and $\epsilon = |R_x/G_x| \approx |R_y/G_y| < 1 $. 
For the large $N$ or the large signal-to-noise ratio $1/\epsilon \gg 1$, $\theta_{xy} \approx \delta_{xy} = \bold{k} \cdot \bold{d}$, which means that we can estimate the wavenumber in the direction of $\bold{d}$ using the cross phase.
In other words, the cross phase $\theta_{xy}(f)$ can be interpreted as the estimation of the local dispersion relation $K(\omega) = \delta_{xy}(\omega)/d \approx \theta_{xy}(\omega)/d$ where $K$ is the local wavenumber along the $\bold{d}$ direction and $\omega = 2 \pi f$ is the measured angular frequency. 
The local poloidal or radial dispersion relation can be obtained with the two-dimensional diagnostics by using the channels separated in the poloidal or radial direction within the correlation length, respectively. 

Using the local dispersion relation, the phase velocity ($\omega / K$) or the group velocity ($\partial \omega / \partial K$) of the fluctuation in the laboratory frame can be obtained.
Since $\omega$ includes the Doppler shift by the plasma flow ($\omega_D = K v$ where $v$ is the plasma flow), the laboratory frame velocity measurements ($v_L$) contain some information about the plasma flow, i.e. $v_L = v_P + v$ where $v_P$ represent the plasma frame measurements. 
If we can assume that $v_P$ is nearly uniform in the measurement region, the spatial variation of $v_L$ would come from the spatial variation of the plasma flow. 
In other words, we can measure the radial or poloidal shear of the plasma flow using $\nabla v \approx \nabla v_L$ in the constant $v_P$ region. 
For the drift wave instability, $v_P$ depends on the pressure gradient and it would be uniform in the constant pressure gradient region. 

\subsection{Local wavenumber-frequency spectrum estimation}

Following the method introduced in~\cite{Beall:1998fx}, the local wavenumber-frequency spectrum $S_L(K,f)$ can be estimated as
\be
S_L(K,f) = \left \langle \left( \frac{X(f)X^*(f) + Y(f)Y^*(f)}{2} \right) \delta_D \left[ \frac{\delta_{xy}(f)}{d} - K \right] \right \rangle
\label{Elfw}
\ee
where $\delta_D[\cdot]$ represents the Dirac delta function and $\delta_{xy}$ can be estimated by the cross phase calculated from each measurement. 
This provides how the fluctuation power is distributed in the wavenumber $K$ and frequency $f$ space by means of a histogram~\cite{Beall:1998fx}.
$S_L(K,f)$ can be more informative than the local dispersion relation, but it is not easy to identify the noise contribution.
The imaging diagnostics could allow a more accurate estimation as used in~\cite{Lee:2016ku}, since multi-pairs of the channels along the same direction can be utilized to perform the ensemble average $\langle S_L(K,f) \rangle_p$ where $\langle \cdot \rangle_p$ represents the average over pairs.

\subsection{Bispectrum and bicoherence estimation}

The nonlinear wave-wave coupling can be identified using the fact that the coupled waves have a well-defined phase relation~\cite{Kim:1979ps}. 
Consider three waves with the amplitudes $G_1$, $G_2$, and $G_3$ and frequencies $f_1$, $f_2$, and $f_3=f_1+f_2$ (satisfying the frequency resonance condition~\cite{Kim:1979ps}) measured by a single channel as $x(t_j)$.
The auto bispectrum is defined as below to investigate the phase coupling among the waves. 
\be
B(f_1, f_2) = \langle X_1 X_2 X^*_3 \rangle 
\ee
The 1st measurement in the ensemble average will be 
\be
X_1^{(1)} X_2^{(1)} X_3^{(1)*} = G_1^{(1)} G_2^{(1)} G_3^{(1)} e^{i(\delta_1^{(1)} + \delta_2^{(1)} - \delta_3^{(1)})} + ...
\ee
where $\delta_1$, $\delta_2$, and $\delta_3$ represent the initial phase of each wave $f_1$, $f_2$, and $f_3$, respectively. 
If three waves are excited independently and the separation time between the measurements is longer than their auto correlation times, their phase difference ($\delta_1 + \delta_2 - \delta_3$) in each measurement will be random.
Then, the $G_1 G_2 G_3$ term will decay with $1/\sqrt{N}$ in the ensemble average. 
On the other hand, if they are nonlinearly coupled, $\delta_1 + \delta_2 - \delta_3 = \mathrm{const}$ and the $G_1 G_2 G_3$ term will remain. 
The bispectrum with $f_1$, $f_2$ and $f_3=f_1+f_2$ is useful to measure the statistical dependence among three waves. 

In practice, the squared bicoherence (the normalized squared bispectrum) is often used to measure the degree of the nonlinear coupling, because we have the finite $N$ and the $1/\sqrt{N}$ decay may not be sufficient for the large amplitudes. 
The squared bicoherence is defined as follows.
\be
b^2(f_1, f_2) = \frac{|\langle X_1 X_2 X^*_3 \rangle |^2}{\langle |X_1X_2|^2 \rangle \langle |X_3|^2 \rangle}
\ee
This value ranges from 0 to 1, meaning the fraction of the power at $f_3$ due to the coupling against the total power at $f_3$. 
$\langle |X_3|^2 \rangle$ represents the total power at $f_3$ and $\frac{|\langle X_1 X_2 X^*_3 \rangle |^2 }{\langle |X_1X_2|^2 \rangle}$ is the power at $f_3$ due to the coupling with $f_1$ and $f_2$~\cite{Kim:1979ps}. 


CORRECTION! Denominator noise contribution. The auto bispectrum might be more preferred than the cross bispectrum (e.g., $\langle X_1 X_2 Y^*_3 \rangle$) since the terms including the independent noise with a short auto correlation time will decay in the ensemble average.
Nonetheless, the cross bispectrum can be useful when we have a particular purpose as described in the next subsection. 
For the cross bispectrum calculation using the spatially separated channels, the channels should be selected carefully that $\bold{d} \cdot \nabla v = 0$ since the different flow velocities would break the frequency resonance condition in the measurements. 

\subsection{Nonlinear energy transfer estimation}

The bispectrum or the squared bicoherence measures the degree of the nonlinear coupling, but the coupling coefficient is a more informative quantity. 
For the simplest system in which the third wave ($f_3$) is just the result of a single three-wave coupling with the others, i.e. $G_3 = Q_{1,2}^{3} G_1 G_2$, the coupling coefficient can be estimated using the bispectrum as $Q_{1,2}^{3} \approx \frac{B^*(f_1,f_2)}{\langle | X_1 X_2 |^2 \rangle}$ for the large signal-to-noise ratio~\cite{Kim:1979ps}.
However, the real system is more complicated and a mode evolves through more than one nonlinear coupling.

For some physical quantity $\phi (z,t)$ at position $z$ and time $t$, we can decompose $\phi (z,t)$ with the wavenumber Fourier coefficients such as $\phi (z,t) = \sum_p \Phi (k_p, t) e^{i k_p z}$ and study its temporal evolution for each $k_p$~\cite{Ritz:1989ci}, or decompose it with the frequency Fourier coefficients such as $\phi (z,t) = \sum_p \Phi (z,\omega_p) e^{i \omega_p t}$ and study its spatial evolution for each $\omega_p$~\cite{Wit:1999jm}. 
For example, the spatial evolution of the fluctuation including many quadratic couplings can be written as 
\be
\frac{\partial \Phi(z, \omega_p) }{\partial z} = \Lambda_{\omega_p}^L \Phi (z, \omega_p) +  \displaystyle\sum_{\substack{p_1 \ge p_2 \\ p=p_1+p_2}}  \Lambda_{\omega_{p_1},\omega_{p_2}}^Q \Phi (z, \omega_{p_1}) \Phi (z, \omega_{p_2}) + ...
\ee 
 where $\Lambda_{\omega_p}^L$ and $\Lambda_{\omega_{p_1},\omega_{p_2}}^Q$ are the linear and quadratic coupling coefficients, respectively. 
 
Next, we need to discretize the equation (12) to reconstruct the following system to utilize the two-point measurements at $z$ and $z+d$~\cite{Ritz:1986gh}. 
\be
Y_p = L_p X_p + \sum_{\substack{p_1 \ge p_2 \\ p=p_1+p_2}} Q_{p_1,p_2}^{p} X_{p_1} X_{p_2} + ...
\ee
where $X_p$ and $Y_p$ represent the measured Fourier coefficients at $z$ and $z+d$, respectively, $L_p$ is the linear transfer function, and $Q_{p_1,p_2}^{p}$ is the quadratic transfer function.
It is to estimate those transfer functions which transform $X_p$ into $Y_p$. 
The linear and quadratic transfer functions for such a system can be calculated with the Millionshchikov hypothesis ($\langle X_{p_1} X_{p_2} X_{p_3}^* X_{p_4}^* \rangle \approx \langle |X_{p_1} X_{p_2}|^2 \rangle$) as shown in~\cite{Ritz:1986gh,Kim:1996jv}.
\begin{eqnarray}
L_p &=& \frac{ \langle Y_p X^*_p\rangle - \displaystyle\sum_{\substack{p_1 \ge p_2 \\ p=p_1+p_2}} \frac{\langle X_{p_1} X_{p_2} X^*_p \rangle \langle X^*_{p_1} X^*_{p_2} Y_p \rangle}{\langle | X_{p_1} X_{p_2} | ^2 \rangle} }{ \langle X_p X^*_p \rangle - \displaystyle\sum_{\substack{p_1 \ge p_2 \\ p=p_1+p_2}} \frac{| \langle X_{p_1} X_{p_2}X^*_p \rangle|^2}{\langle |X_{p_1} X_{p_2}|^2 \rangle} } \\
Q_{p_1,p_2}^{p} &=& \frac{\langle X^*_{p_1} X^*_{p_2} Y_p \rangle - L_p \langle X^*_{p_1} X^*_{p_2} X_p \rangle}{\langle | X_{p_1} X_{p_2} |^2 \rangle}
\end{eqnarray}
Or, we can directly solve the following matrix equation for $\bold{H}$ with the large number of ensembles ($N \gg$ the number of unknowns in $\bold{H}$)~\cite{Wit:1999jm}.
\begin{figure}
\includegraphics[keepaspectratio,width=0.9\textwidth]{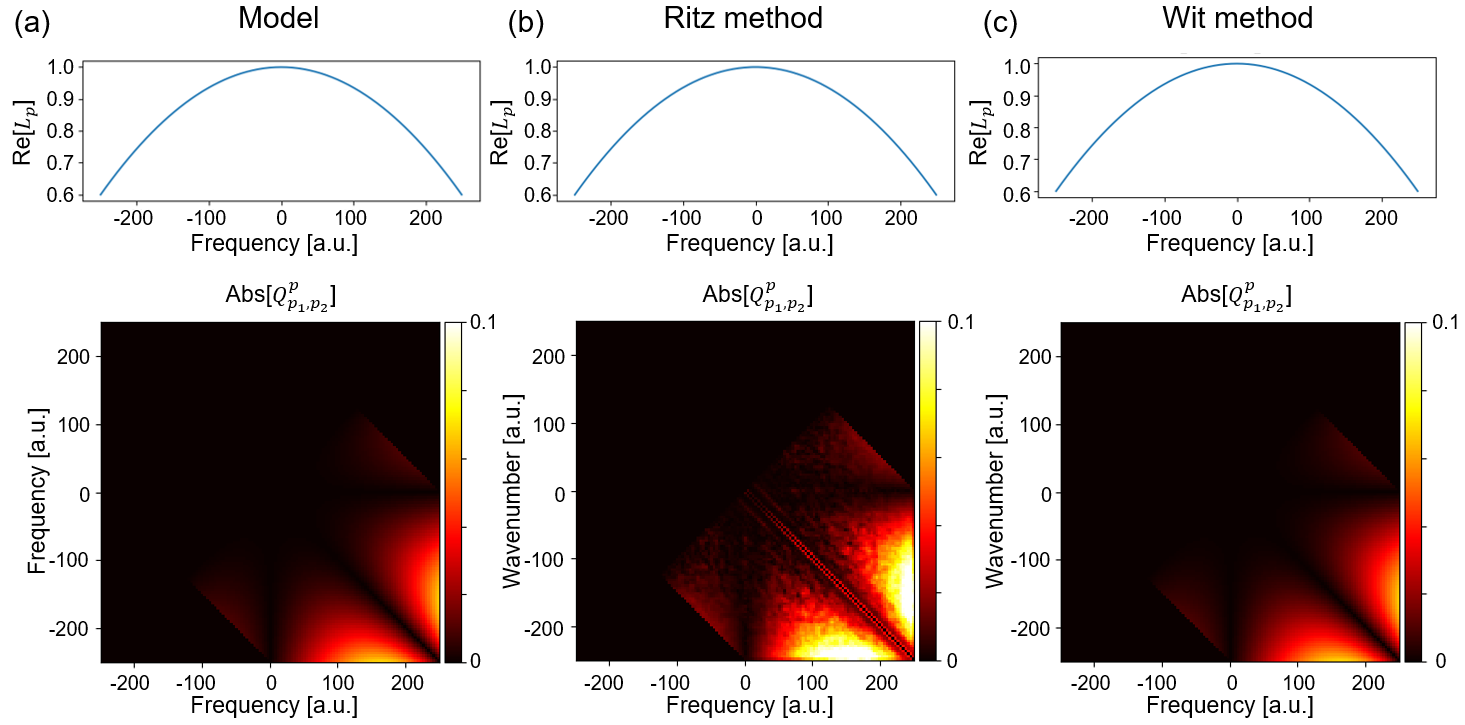}
\centering
\caption{(Color online) (a) The real part of $L_p$ and $|Q_{p_1,p_2}^p|$ from the model equation. Estimation results using (b) the Ritz and (c) Wit method.}
\label{Ftrw}
\end{figure}
\be
\bold{Y} = \bold{X} \bold{H}
\ee
where
\begin{eqnarray}
\bold{Y} = 
	\begin{bmatrix}
	Y_p^{(1)} \\
	Y_p^{(2)} \\
	\vdots \\
	Y_p^{(N)} \\
	\end{bmatrix}
~\bold{X} = 
	\begin{bmatrix}
	X_p^{(1)}  & X_{p_1}^{(1)} X_{p-p_1}^{(1)} & X_{p_2}^{(1)} X_{p-p_2}^{(1)} & ... \\
	X_p^{(2)}  & X_{p_1}^{(2)} X_{p-p_1}^{(2)} & X_{p_2}^{(2)} X_{p-p_2}^{(2)} & ... \\
	\vdots & \vdots & \vdots & ... \\
	X_p^{(N)} & X_{p_1}^{(N)} X_{p-p_1}^{(N)} & X_{p_2}^{(N)} X_{p-p_2}^{(N)} & ... \\
	\end{bmatrix}
~\bold{H} = 
	\begin{bmatrix}
	L_p \\
	Q_{p_1,p-p_1}^p\\
	Q_{p_2,p-p_2}^p\\
	\vdots \\
	\end{bmatrix}
\end{eqnarray}
Verification of these two methods (the Ritz~\cite{Ritz:1986gh,Kim:1996jv} and Wit~\cite{Wit:1999jm} methods) using the model $L_p$ and $Q_{p_1,p_2}^p$ are shown in figure~\ref{Ftrw}. 
The following model equations are used for $L_p$ and $Q_{p_1,p_2}^p$ to generate $Y_p$ from some given $X_p$ data~\cite{Ritz:1986gh}.
\begin{eqnarray}
L_p &=& 1.0 - 0.4 \frac{p}{p_{Nyq}^2} + i0.8 \frac{p}{p_{Nyq}} \\
Q_{p_1,p_2}^p &=& \frac{i}{5 p_{Nyq}^4} \frac{p_1 p_2 (p_2^2 - p_1^2)}{1 + p^2/p_{Nyq}^2}
\end{eqnarray}
where $p_{Nyq}$ is the index of the Nyquist frequency (250 in this case). 

%

Once the transfer functions ($L_p$ and $Q_{p_1,p_2}^p$) are estimated with $X_p$ and $Y_p$ either using the Ritz (with the Millionshchikov hypothesis) or Wit method, we can calculate a spatial linear growth rate $\gamma_p$ and a spatial nonlinear energy transfer rate $T_p$ of the spectral power $P_p = \langle X_p X_p^* \rangle$ whose evolution is described as
\be
\frac{\partial P_p}{\partial z} \approx \frac{\langle Y_p Y_p^* \rangle - \langle X_p X_p^* \rangle}{d} = \gamma_k P_k + T_k
\ee
where~\cite{Kim:1996jv}
\begin{eqnarray}
\gamma_p &\approx& \frac{|L_p|^2 - 1}{d} \\
T_p &\approx& 2~\mathrm{Re} \left[ L_p^* \sum_{\substack{p_1 \ge p_2 \\ p=p_1+p_2}} \frac{Q_{p1,p2}^p \langle X_{p1} X_{p2} X_p^* \rangle}{d}   \right] + \sum_{\substack{p_1 \ge p_2 \\ p=p_1+p_2}}\sum_{\substack{p_3 \ge p_4 \\ p=p_3+p_4}} \frac{Q_{p1,p2}^p Q_{p3,p4}^{p*} \langle X_{p1} X_{p2} X_{p3}^* X_{p4}^* \rangle}{d}
\end{eqnarray}


\section{Practical application examples}

\subsection{Frequency spectrum measurement}

\begin{figure}
\includegraphics[keepaspectratio,width=0.45\textwidth]{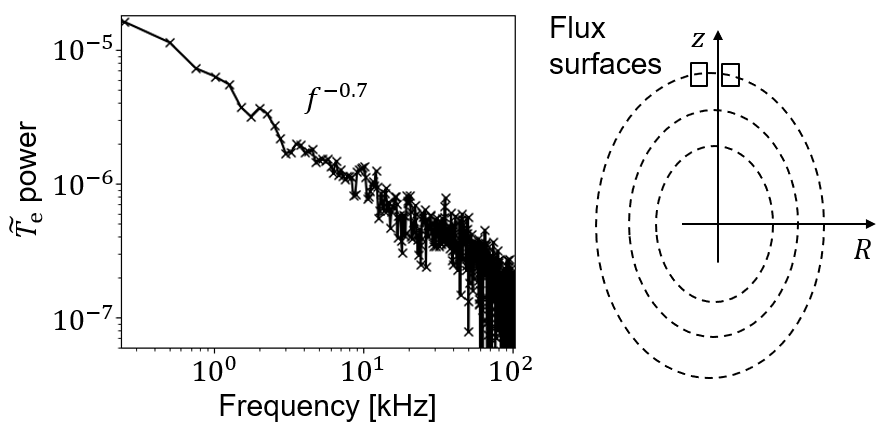}
\centering
\caption{The power spectrum of the event size exhibits the power-law behavior. Two adjacent channels (square boxes) on the same flux surface are used to reduce the noise contribution in the power spectrum.}
\label{Fcpr}
\end{figure}

An accurate measurement of the frequency spectrum is not only the basis of all spectral analyses but it is an important element to validate the transport model.
For example, the non-diffusive transport model based on the self-organized criticality predicts the power-law behavior of the avalanche event size power spectrum $S(f)\propto f^{-\alpha}$ where $\alpha = 1$ for the ideal case and $0 < \alpha <1$ for the case including a subdominant diffusive transport~\cite{Newman:2015jp, Hahm:2018dm}.
In the KSTAR plasma in which the magnetohydrodynamic instabilities are quiescent, the avalanche-like events are observed.
They produce the $m=0$ electron temperature bump ($\delta T_\mathrm{e} > 0$) and void ($\delta T_\mathrm{e} <0$) propagating radially in opposite directions. 
Two ECEI channels on the same flux surface were able to measure the power spectrum of the $\delta T_\mathrm{e}$ size of the avalanche-like events via the cross power. 
Figure~\ref{Fcpr} shows the result which exhibits the power-law behavior $S(f) \propto f^{-0.7}$ as expected from the self-organized criticality theory.
The accurate power spectrum measurement could identify the non-diffusive avalanche-like characteristics of the electron heat transport in this plasma~\cite{ChoiNF2019}. 

\subsection{Flow shear measurement}

\begin{figure}
\includegraphics[keepaspectratio,width=0.4\textwidth]{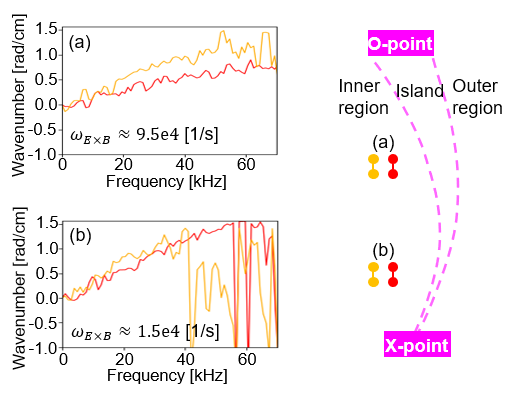}
\centering
\caption{The cross phase measurements in different regions outside the magnetic island.}
\label{Fcph}
\end{figure}

The radial shear of the plasma flow is important since it can suppress the mode whose growth rate is smaller than the flow shear [reference]. 
As discussed in Section 2.2, the flow shear can be estimated using local dispersion measurements with the constant $v_0$ assumption.
Figure~\ref{Fcph} shows four local dispersion measurements at different locations in the inner region of the magnetic island close to the midplane ($z = \pm 5$~cm).
Each local dispersion is obtained from the cross phase ($\delta_{xy}(f) = K(f) d$) between vertically adjacent ECEI channels. 
Two radially adjacent measurements of the local dispersion are used to estimate the poloidal flow shear near the O-point (figure~\ref{Fcph}(a)) and X-point (figure~\ref{Fcph}(b)).
The increasing flow shear towards the O-point is identified using multi-location measurements~\cite{Choi:2017ez}, which shows the strength of the two-dimensional diagnostics. 
The observed radial shear of the flow is attributed to the $E \times B$ flow perturbation by the magnetic island. 
This finding explained that the fluctuation power becomes weaker as it goes far from X-point~\cite{Choi:2017ez, Kwon:2018bi, Fang:2019fz}. 

\subsection{Nonlinear interaction measurement (preliminary)}

\section{Summary}

Some useful spectral methods for the two-dimensional diagnostics are introduced. 
The strengths of the two-dimensional diagnostics in investigating the frequency spectrum, the local dispersion relation, the flow shear, and the nonlinear interaction have been demonstrated with practical applications. 

\section{Appendix}

\subsection{fluctana}

The Python code package named as ``fluctana'' has been developed to provide an easy access and analysis of the various fluctuation data of the KSTAR tokamak. 
It includes all the spectral methods introduced in this paper as well as the statistical methods to calculate the higher order moments, the Hurst exponent, the Jensen-Shannon complexity and the normalized Shannon entropy, and the transfer entropy. 
The code and simple tutorials are available via the GitHub repository~\url{https://github.com/minjunJchoi/fluctana}. 

\bibliographystyle{ieeetr}

\end{document}